\documentclass[12pt,a4paper]{article}
\usepackage[utf8]{inputenc}
\usepackage[T1]{fontenc}
\usepackage{graphicx}

\begin{document}
\textwidth=135mm
 \textheight=200mm
\begin{center}
{\bfseries Cooling of neutron stars and hybrid stars with a \\stiff hadronic EoS}
\vskip 5mm
H. Grigorian$^{\ast,\dag}$\footnote{Email: hovik.grigorian@gmail.com},
D. Blaschke$^{\ddag,\star}$,
D. N. Voskresensky$^{\S}$
\vskip 5mm
{\small {\it $^\ast$ Laboratory for Information Technologies,
JINR Dubna, 141980 Dubna, Russia}} \\
{\small {\it $^\dag$ Department of Physics, Yerevan State University, 0025 Yerevan, Armenia}} \\
{\small {\it $\ddag$ Instytut Fizyki Teoretycznej, Uniwersytet Wroc\l{}awski,
50-204 Wroc\l{}aw, Poland}}\\
{\small {\it $^\star$ Bogoliubov Laboratory for Theoretical Physics,
JINR, 141980 Dubna, Russia}} \\
{\small {\it $\S$ National Research Nuclear University (MEPhI),
115409 Moscow, Russia}}
\end{center}
\vskip 5mm \centerline{\bf Abstract} Within the "nuclear medium
cooling" scenario of neutron stars all reliably known temperature
- age data, including those of the central compact objects in the
supernova remnants of Cassiopeia A and XMMU-J1732, can be
comfortably explained by a set of cooling curves obtained by
variation of the star mass within the range of typical observed
masses. The recent measurements of the high masses of the pulsars PSR
J1614-2230 and PSR J0348-0432 on the one hand, and of the low 
masses for PSR J0737-3039B and the companion of
PSR J1756-2251 on the other, provide independent proof for the 
existence of neutron stars with masses in a broad range from 
$\sim 1.2$ to 2 $M_\odot$.  
The values $M>2~M_{\odot}$  call for sufficiently stiff equations of
state for neutron star matter. We investigate the response of the
set of neutron star cooling curves to a stiffening of the nuclear
equation of state so that maximum masses of about $2.4~M_\odot$
would be accessible and to a deconfinement phase transition from
such stiff nuclear matter in the outer core to colour
superconducting quark matter in the inner core. Without a
readjustment of cooling inputs the mass range required to cover
all cooling data for the stiff DD2 equation of state should
include masses of $2.426~M_\odot$ for describing the fast cooling
of CasA while the existence of a quark matter core accelerates the
cooling so that CasA cooling data are described with a hybrid star
of mass $1.674~M_\odot$. \vskip 10mm

\section{\label{sec:intro}Introduction}
The cooling of compact stars (CS) is an observable phenomenon
which is governed by the interplay of structure and composition
(viz. the equation of state (EoS)) of CS with the transport
properties and neutrino emissivities of the matter they are made
of. It therefore allows, in principle, to explore the otherwise
inaccessible physics of neutron star interiors. Until recently the
complex situation with many poorly known parameters in the theory
of CS cooling allowed for many possible scenarios due to the fact
that the observational data on masses and radii as well as
temperature and age of CS were not sufficiently constraining.

Now the situation has tremendously improved with the observation
of the segment of a cooling curve for the central compact object
in the remnant of the historical supernova Cassiopeia A
\cite{Tananbaum:1999kx,Ashworth:1980vn}, thus with known age,
temperature and rate of cooling followed over the past 13 years
since its discovery
\cite{Ho:2009fk,Heinke:2010xy,Elshamouty:2013nfa}. In principle,
the observed spectra and distance allow even for a rough
constraint on mass and radius of the cooling CS
\cite{Ho:2009fk,Yakovlev:2010}. These data require a fast cooling
process in the CS interior which becomes apparent in the photon
luminosity at the surface of the star after 300 yr. On the other
hand, the CS cooling model must also explain that XMMU
J173203.3-344518 \cite{Klochkov}, another compact object in a
supernova remnant, being  hotter and older than CasA, at an age
between 10 and 40 kyr.  The solution of the puzzle might be
connected with a strong medium dependence of cooling inputs, as
provided by the density dependent
medium modifications of the nucleon-nucleon interaction caused by
the softening of the pion degree of freedom with  the density, and
by the density dependent superfluid pairing gaps, see
\cite{Voskresensky:2001fd} for details. The key idea that the
cooling of various sources should be essentially different due to
the difference in their masses was formulated long ago
\cite{Voskresensky:1986af}, when still there existed the opinion
that all masses of neutron stars should be approximately fixed
closely to the value $1.4~M_{\odot}$. The recent measurements of
the masses of the pulsars PSR J1614-2230 \cite{Demorest}, PSR
J0348-0432 \cite{Antoniadis} and J00737-3039B \cite{Kramer} and
of the companion of PSR J1756-2251 \cite{Faulkner} provide the proof
for the existence of CS with masses varied in a broad range, at
least from $\sim 1.2$ to $2~M_\odot$.

First, influence of in-medium effects based on assumption
\cite{Voskresensky:1986af} on the neutron star cooling was
demonstrated in \cite{Schaab:1996gd} within  various exploited
EoS. The "nuclear medium cooling" scenario subsequently developed
 in \cite{Blaschke:2004vq,Blaschke:2011gc} provides a successful
description of all known cooling data for neutron stars with low
magnetic fields. Recently, it has even been improved
\cite{Blaschke:2013vma} in order to comply with the constraint
that the EoS used for calculating the CS sequence should reach a
maximum mass in the range $M=2.01\pm 0.04~M_\odot$ as measured for
PSR J0348+0432 \cite{Antoniadis}, see also \cite{Demorest}. The
most efficient processes within the nuclear medium cooling
scenario are the medium modified Urca (MMU) process, e.g. $nn\to
npe\bar{\nu}$, and the pair-breaking-formation processes (PBF).
The latter processes, $N\to N_{\rm pair}\nu\bar{\nu}$, $N=n$ or $p$,
are enhanced owing to their one-nucleon nature
\cite{Voskresensky:1987hm,Kolomeitsev:2008mc}, despite they are
allowed only in the presence of the nucleon pairing.  The direct
Urca reaction, $n\to npe\bar{\nu}$, is forbidden at least for
$M<1.5~M_{\odot}$ (so called "strong" DU constraint), see
\cite{Klahn:2006ir}.

Still, the EoS used in \cite{Blaschke:2013vma} might be not
sufficiently constrained. A future measurement of  radii of CS
might require stiffer hadronic EoS, which would entail a
restructuring of the star and modification of its cooling
characteristics. Recent radius determinations from timing
residuals suggest the larger radii and thus stiffer hadronic EoS
\cite{Bogdanov:2012md}. Moreover, for stiffer hadronic EoS, a
deconfinement transition in the CS interior is not excluded
\cite{Alford:2006vz,Klahn,Klahn2}. Also, more massive objects than
those have been carefully measured in \cite{Demorest,Antoniadis}
are not excluded. For example there are some although indirect
indications \cite{Romani} that the mass of the black-widow pulsar
PSR J1311-3430 may even reach $M = 2.7~M_{\odot}$. Incorporating
systematic light-curve differences the authors estimated that the mass 
should be at least $M>2.1~M_{\odot}$.

Therefore, in the present contribution we shall explore these
actual EoS aspects. Here we compute the cooling curves within our
nuclear medium cooling scenario exploiting the stiff DD2 EoS
\cite{Typel:2009sy}. Additionally, as an alternative to the purely
hadron scenario, we incorporate a possibility of the deconfinement
phase transition from such stiff EoS of the nuclear matter in the
outer core to colour superconducting quark matter in the inner
core.

\section{\label{sec:EoS}Stiff hadronic EoS and high-mass hybrids}

The baseline EoS for our previous work was the APR based fit
formula provided by Heiselberg and Hjorth-Jensen
\cite{Heiselberg:1999fe}
 \begin{equation}\label{EN}
 E_N=u n_0\left[m_N+ e_{\rm B} u\frac{2+\delta -u}{1+\delta  u} +
 a_{\rm sym}u^{0.6}(1-2x_p)^2\right]\,,
  \end{equation}
where $u=n/n_0$, with $n_0=0.16$ fm$^{-3}$ the nuclear saturation density,
$e_{\rm B}\simeq -15.8$ MeV is the nuclear
binding energy per nucleon, $a_{\rm sym}\simeq 32$ MeV is the
symmetry energy coefficient and we chose $\delta =0.2$.
We have recently improved this EoS by invoking an excluded volume
for nucleons related to their quark substructure and the Pauli blocking
between nucleons due to quark exchange forces. This was accomplished
by the replacement
\begin{equation}
u\longrightarrow u\left[1- n_0v_0 {\rm e}^{-(\beta/u)^\sigma} \right]^{-1}~,
\end{equation}
where $v_0=0.125$ fm$^3$ stands for the effective excluded volume,
$\beta=6$ and $\sigma=4$ \cite{Blaschke:2013vma}.

\begin{figure}[!thb]
   \centering
   \includegraphics[width=0.95\textwidth,keepaspectratio=true]{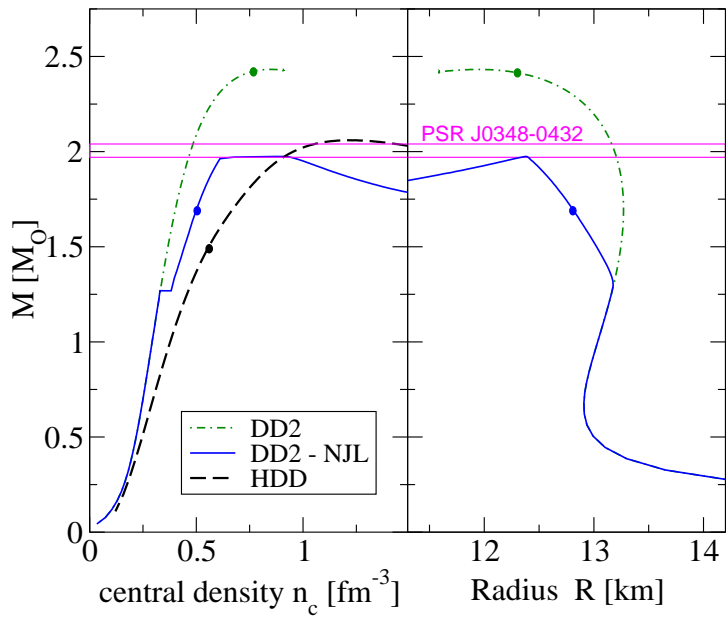}
     \caption{Mass vs. central baryon density (left) and vs. radius
     (right) for the stiff DD2  hadronic EoS (dash-dotted lines)
     and for the hybrid EoS with a deconfinement transition to superconducting
     NJL quark matter (solid lines). For comparison, the softer hadronic EoS model HDD 
     \cite{Blaschke:2013vma}
     is shown in the left panel (dashed line). All three EoS fulfil
     the constraint for the maximum mass of $1.97~M_\odot$.
     The dots indicate the configurations for which the cooling curve
      would describe the observational data from CasA.}
   \label{Fig:MR}
 \end{figure}

The sequence of CS resulting from integrating the
Tolman--Oppenheimer--Volkoff equation is shown in the left panel
of Fig.~\ref{Fig:MR}, denoted as "HDD" and it serves as the
reference point for the present study. The black dot on that curve
at the mass $1.498~M_\odot$ (for the pairing gaps computed
following model I, see  \cite{Blaschke:2013vma}) denotes the star
configuration which would describe the cooling curve that fits
best the CasA cooling data.

 There are indications from a recent radius determination for the nearest
 millisecond pulsar PSR J0437-4715 \cite{Bogdanov:2012md} that
a stiffer EoS is required to support (at $2\sigma$ confidence)
radii  $\geq 13$ km  in the mass segment between 1.5 and 1.8
$M_\odot$. The density dependent relativistic mean-field EoS of
Ref.~\cite{Typel:1999yq} with the well calibrated DD2
parametrization \cite{Typel:2009sy} meets such requirements, see
the dash-dotted line in the right panel of Fig.~\ref{Fig:MR}. It
fulfills all standard constraints for symmetric nuclear matter
around saturation density and from nuclear structure. It has a
density dependent symmetry energy in perfect agreement with the
recent constraint by Danielewicz and Lee
\cite{Danielewicz:2013upa} and with ab-initio calculations for
pure neutron matter \cite{Hebeler:2010jx}. The DU threshold is not
reached within the DD2 EoS that is also a good feature, since all
the stars with the density in the center above the DU threshold
value cool down so rapidly that become invisible in the soft $X$
rays. Note however that due to the stiffness of this EoS, it does
not fulfil the Danielewicz "flow constraint"
\cite{Danielewicz:2002pu} for densities above $2n_0$. This is the
price we pay for a possibility to substantially increase the
maximum neutron star mass and the radii of CS.

This DD2 EoS will be used by us as a benchmark for a stiff
hadronic EoS. It is plain from inspecting the left panel of
Fig.~\ref{Fig:MR} that stiffening the EoS does not only entail
larger CS radii but at the same time lower densities of the
neutron star interior. As a consequence, a slower cooling than for
stars of the HDD sequence is expected at the same mass. In other
words, in order to cover the same set of cooling data with a
stiffer EoS the range of masses attributed to the set of cooling
curves shall be shifted to higher values.

A sufficiently stiff hadronic EoS like DD2 paves the way for
exploring scenarios with phase transitions to exotic forms of
matter in the CS interior like hyperons \cite{Lastowiecki:2011hh}
and/or quark matter, see  Refs. \cite{Alford:2006vz,Klahn,Klahn2}
for recent examples which despite a softening due to the phase
transition meet the constraint of the $2~M_\odot$ pulsar mass
measurement \cite{Antoniadis}, see the solid lines in
Fig.~\ref{Fig:MR}. In those examples the quark matter EoS is
described by a colour superconducting two-flavor NJL model in the
2SC phase with a stiffening due to a vector mean field. We shall
now discuss the results for CS cooling obtained with these EoS.

 \section{\label{sec:Cooling}Cooling model of CS}

Since we would like to isolate the effects of changing just the
EoS on the cooling behaviour, we adopt here all cooling inputs
such as  the neutrino emissivities, specific heat, crust
properties, etc., from our earlier works performed on the basis of
the HHJ EoS \cite{Blaschke:2011gc} and HDD EoS
\cite{Blaschke:2013vma} for the hadronic matter. The pairing gaps
will be taken following the model II. Of key importance is
that we will use here the very same   density dependence of the
effective pion gap $\widetilde{\omega} (n)$ as in our  other
previous works, e.g., see  Fig.~1 of \cite{Blaschke:2013vma}. 
To be specific we assume the pion condensation to appear for $n>3n_0$
thus exploiting  the curves 1a+2+3.

The heat conductivity is treated in the same simplified way as in
\cite{Blaschke:2011gc}. There, we used an additional suppressing
pre-factor $\zeta_{\kappa}$ to show the qualitative effect. With the
gaps from the model II, the best description of the CasA data was
achieved with the mass $M\geq 1.73~M_{\odot}$ at the parameter
$\zeta_{\kappa}\leq 0.015$.
Since our aim here is just to demonstrate the qualitative effect 
of a stiffening of the EoS on the cooling we will follow the same 
simplified procedure.

In order to describe the possibility of hybrid star configurations
we adopt the cooling of the quark core following the lines of
\cite{Grigorian:2004jq}. There the quark core cooling is described
by adopting a small but nonvanishing density dependent pairing gap
(X-gap) for those quarks with the otherwise in the 2SC phase unpaired colour.

\section{\label{sec:Results}Results}

\begin{figure}[!thb]
   \centering
   \includegraphics[width=0.95\textwidth]{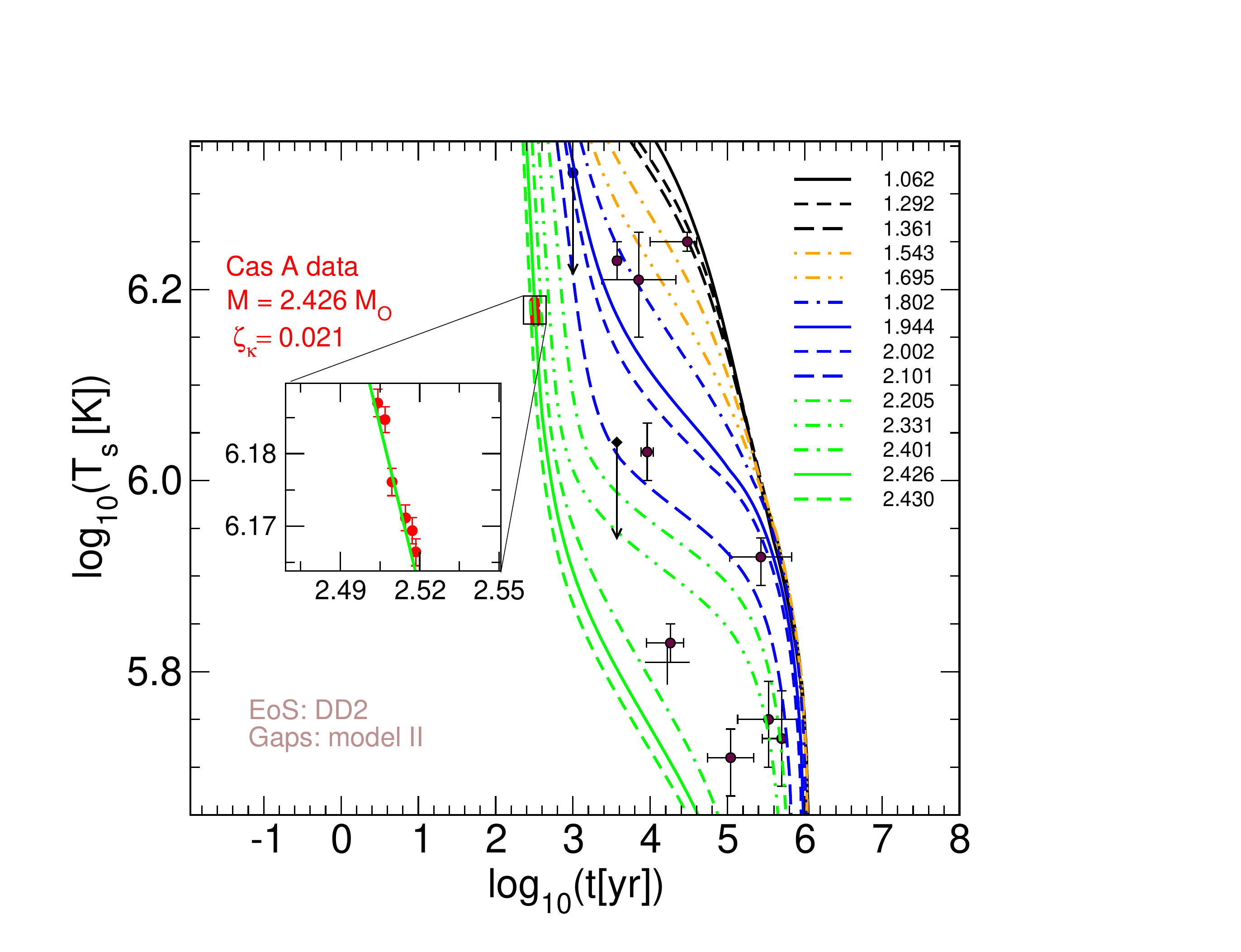}
      \caption{Cooling curves for a neutron star sequence according to the stiff 
      hadronic DD2 EoS;
      $T_s$ is the surface temperature, $t$ is the CS age.
      The color coding corresponds to different mass ranges: $M=1.0 - 1.5~M_\odot$ 
      (black), $M=1.5 - 1.8~M_\odot$ (orange), $M=1.8 - 2.2~M_\odot$ (blue) and
      $M=2.2 - 2.5~M_\odot$ (green). Cooling data for CasA are
      explained with a heavy neutron star of $M=2.426~M_\odot$. 
      To recover an appropriate slope of the  curve requires a pre-factor
      $\zeta_{\kappa}=0.021$.}
   \label{Fig:Cool1}
 \end{figure}

For the purely hadronic scenario the resulting cooling curves are shown in 
Fig.~\ref{Fig:Cool1}.
As expected, the stiffer DD2 hadronic EoS leads to a flatter
density profile and therefore to weaker cooling activity when
compared to the relatively soft HHJ and HDD EoS, provided the same
effective pion gap $\widetilde{\omega} (n)$ is used. Consequently
a rather large CS mass range is required in order to cover the
full set of cooling data. It is a remarkable and nontrivial fact
that the description of all cooling data after this change of the
EoS is possible without changing any of the formerly adjusted
cooling inputs! Note that the mass $M=2.426~M_\odot$ fitted for
the CasA cooling data is in the range given by the original
analysis of Ho and Heinke \cite{Ho:2009fk}. The fit of the slope,
as indicated in the inset of the figure, requires the choice of
the thermal conductivity suppression pre-factor
$\zeta_{\kappa}=0.021$. The cooling of the hot source XMMU-J1732
is explained by a neutron star with the mass $1.29~M_{\odot}$
being again larger than the ($\simeq 1~M_{\odot}$) which was required
with the HHJ and HDD EoS.

\begin{figure}[!hbt]
   \centering
   \includegraphics[width=0.95\textwidth]{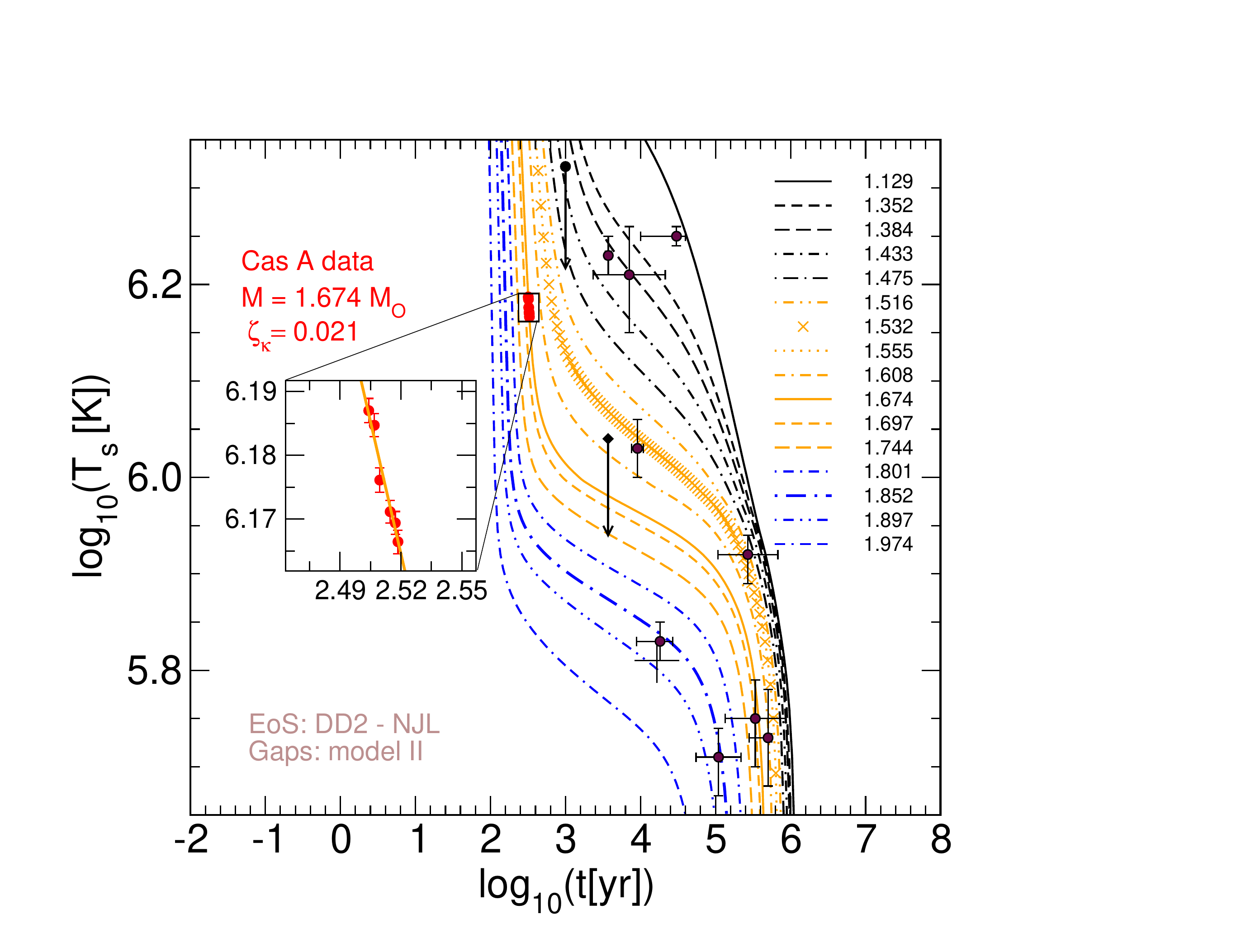}
     \caption{Cooling curves for a CS sequence according to the DD2 - NJL hybrid EoS.
      The color coding is as in Fig.~\ref{Fig:Cool1}.
      Cooling data for CasA are explained with a quark core hybrid star of 
      $M=1.674~M_\odot$.
   To recover an appropriate slope of the  curve requires a pre-factor 
   $\zeta_{\kappa}=0.021$.}
   \label{Fig:Cool2}
 \end{figure}

Turning our attention to the scenario of a hybrid EoS, we observe
that the presence of a quark core leads to an acceleration of the
cooling. The full set of cooling data can again be described (see
Fig.~\ref{Fig:Cool2}) but this time by varying the star masses
within a range much narrower than for the stiff, purely hadronic
DD2 sequence. The cooling data for CasA are now explained with
quark core hybrid star of $M=1.674~M_\odot$. An appropriate slop
of the  curve requires the same  pre-factor
$\zeta_{\kappa}=0.021$, as has been used in the purely hadronic
scenario, see Fig. \ref{Fig:Cool1}.

\section{\label{sec:Remarks}Concluding remarks}

We have demonstrated that  the presently known cooling data on
Cassiopeia A and the hot source  XMMU-J1732, as well as other cooling
data,  can be appropriately described within our nuclear medium
cooling scenario, under the assumption that different sources have
different masses, either by purely hadronic or by
hybrid star configurations. Large values of the compact star radii
and the maximum mass, as it might be motivated by observations,
are compatible with our nuclear medium cooling scenario provided
one uses a stiff EoS. Here we demonstrated it with the DD2 EoS.
For computing the cooling curves we exploited the very same
parameter set of the purely hadronic model as in \cite{Blaschke:2011gc}. 
If we allowed for a decrease of the effective pion gap with increasing
density we could diminish the resulting values of the mass of
the neutron star in Cassiopeia A. This will be demonstrated elsewhere.

\vskip 10mm \centerline{\bf Acknowledgment}\vskip 5mm

This work was supported by the NCN ``Maestro'' programme under
contract number UMO-2011/02/A/ST2/00306 and by  the Ministry of
Education and Science of the Russian Federation, in framework of
"Basic part". H.G. acknowledges support by the Bogoliubov-Infeld
programme  for exchange between JINR Dubna and Polish Institutes.
The authors are grateful for support from the COST Action MP1304
"NewCompStar" for their networking and collaboration activities.

\end{document}